\documentclass[twocolumn, prl]{revtex4}
\usepackage{graphicx}
\begin{document}
\title{Motif Conservation Laws for the Configuration Model}
\author{Anatol E. Wegner}
\affiliation{Max Planck Institute for Mathematics in the Sciences, Inselstr. 22, Leipzig-Germany}
\email{wegner@mis.mpg.de}
\begin{abstract}
The observation that some subgraphs, called motifs, appear more often in real networks than in their randomized counterparts has attracted much attention in the scientific community. In the prevalent approach the detection of motifs is based on comparing subgraph counts in a network with their  counterparts in the configuration model with the same degree distribution as the network. In this short note we derive conservation laws that relate motif counts in the configuration model and discuss their consequences.
\end{abstract}
\maketitle
\section{Introduction}
Motif identification \cite{motif,super} has become a widely used method in network analysis. The prevalent approach to motif analysis is due to Milo et al.  and is based on comparing subgraph counts of motifs in the network with their counterparts in a null model that preserves certain features of the network. The most widely used null model is the configuration model \cite{motif,newmanrandom,milouniform} with the same degree distribution as the network. In this short note we present some simple conservation laws relating motif counts that follow directly from the conservation of the degree sequence. Some conserved quantities were given by Milo et. al \cite{super} before and correlations between motif counts have also been investigated in \cite{motifset}. The conservation laws we present here directly relate motif counts and account for the correlations observed between motifs \cite{motifset} and the general structure of motif significance profiles that have been used to categorize networks \cite{super}.

\section{The configuration model}

The configuration model for directed graphs on $n$ nodes \cite{newmanrandom} is based on assigning to each node a specific in, out and mutual degree ($I_i$, $O_i$ and $M_i$, $i=1,2,\ldots,n$) and assigning equal probability to each possible graph configuration with the given degree sequence. Whether to include the mutual degree in the construction or not is a matter of choice but is in general done  when detecting motifs \cite{motif}. Graphs with self edges, parallel edges and additional mutual edges that arise during the randomization process are in general discarded from the ensemble. In the case of undirected graphs one can simply consider all edges to be mutual edges. If additional mutual edges and/or parallel edges are allowed to form during the randomization process the conservation laws we derive hold only approximately. However, the expected number of such edges in general is small (i.e. O(1)). Algorithms for sampling the configuration model are reviewed in \cite{milouniform}. 
\section{Results}
\subsection{Definitions and Conventions}

The conservation laws we present are based on the distinction between subgraphs and induced subgraphs and the observation that subgraph counts of V-shaped motifs are preserved in the configuration model. 

A graph $H=(V(H),E(H))$ is called a subgraph of $G=(V(G),E(G))$ whenever $V(H)\subseteq V(G)$ and $E(H)\subseteq E(G)$. A subgraph is said to be \emph{induced} iff it contains all edges $xy\in E(G)$ such that $x,y\in V(H)$. In the literature on network motifs the word 'subgraph' in general refers to an induced subgraph and most motif detection algorithms are based on counting induced subgraphs \cite{motif,super}. 
\begin{figure}[h]
\begin{center}
\includegraphics[scale=0.3]{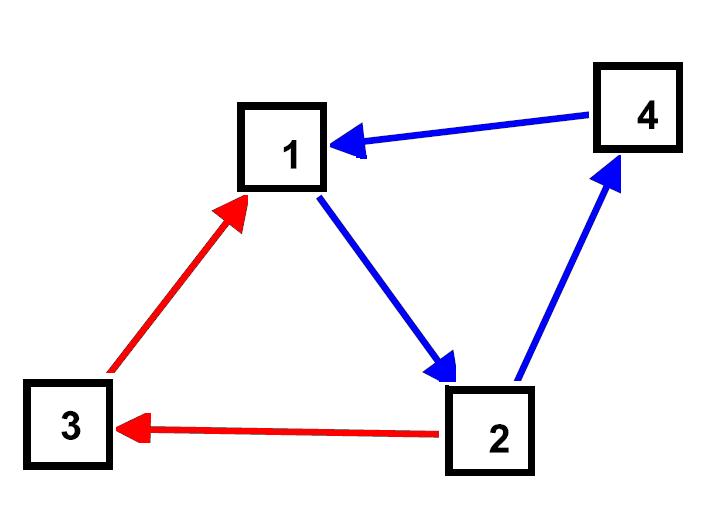}
\label{Fig.1}
\caption {Edges in blue form are an induced subgraph as it contains all edges between nodes 1,2 and 4.While the red edges are not an induced subgraph as they do not contain all edges between nodes 1,2 and 3.}
\end{center} 
\end{figure}

In this paper we consider the configuration model where the mutual degree is conserved therefore we consider mutual edges not as combination of two edges but instead as edges of a different type. This coincides with the convention used to count motifs in \cite{motif, super}. Consequently, in the case of directed 3 node motifs [Fig.2], motif 3 is not considered to be a subgraph of motif 4, neither is 8 a subgraph of 12, etc. If one considers mutual edges to be combinations of two directed edges the counting convention has to be modified accordingly. On the other hand, the conservation laws arising from such a counting convention can be shown to be linear combinations of the ones we derive here. Obviously, the number of conservation laws would decrease if the mutual degree sequence is not conserved. 
\begin{figure}[h]
\begin{center}
\includegraphics[scale=0.45]{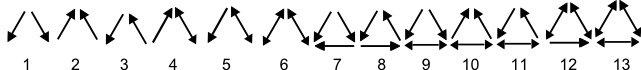}
\label{Fig.2}
\caption {The 13 directed 3-node motifs}
\end{center} 
\end{figure}
\subsection{Conservation laws for directed 3-node motifs}
For a graph $G$ on $n$ nodes with given in, out and mutual degree sequences ($I_i$, $O_i$ and $M_i$, i=1,2,\ldots,n ) the subgraph counts of the V-shaped motifs are entirely determined by moments of the degree sequences and are given by:
\begin{equation}
N_1=\sum_{i=1}^{n}\left(
\begin{array}{c}
O_i\\
2\end{array}
\right)
\end{equation}
\begin{equation}
N_2=\sum_{i=1}^{n}\left(
\begin{array}{c}
I_i\\
2\end{array}
\right)
\end{equation}
\begin{equation}
N_3=\sum_{i=1}^{n} O_i I_i
\end{equation}
\begin{equation}
N_4=\sum_{i=1}^{n} I_i M_i
\end{equation}
\begin{equation}
N_5=\sum_{i=1}^{n}O_i M_i
\end{equation}
\begin{equation}
N_6=\sum_{i=1}^{n}\left(
\begin{array}{c}
M_i\\
2\end{array}
\right)
\end{equation}

It follows that the subgraph counts of the V-shaped triads are conserved in the configuration model as they are functions of the degree sequences only. Since a subgraph is either an induced subgraph or not, the subgraph count of a given motif is simply the sum of the subgraphs which are induced subgraphs and the ones that are not. Moreover, according to our convention V-shaped subgraphs that are not induced subgraphs have to be contained in some triangle shaped induced subgraph. Again because of the convention all triangle shaped subgraphs are induced subgraphs. Since every triangle shaped subgraph contains a certain number of copies of V-shaped motifs as subgraphs, we get the following conservation laws where $N_i$ denotes the subgraph count of motif $i$ and $n_i$  its induced subgraph count:
\begin{equation}
N_1=n_1+n_7+n_9
\end{equation}
\begin{equation}
N_2=n_2+n_7+n_{10}
\end{equation}
\begin{equation}
N_3=n_3+n_7+3n_8+n_{11}
\end{equation}
\begin{equation}
N_4=n_4+2n_9+n_{11}+n_{12}
\end{equation}
\begin{equation}
N_5=n_5+2n_{10}+n_{11}+n_{12}
\end{equation}
\begin{equation}
N_6=n_6+n_{12}+3n_{13}
\end{equation}

These conservation laws show that the statistics of the V-shaped motifs are fully determined by the statistics of the triangle shaped motifs. In the supporting material of \cite{super} it was shown that there are 9 conserved quantities for the sixteen 3-node motif counts (including the single edged and empty motifs). The conservation laws are also closely related to the reactions proposed in \cite{super,motifset} as they represent analogues of mass conservation laws for these reactions. 

\subsection{4 node motifs}
In the case of undirected 4-node motifs there is one analogous conservation law for the 3-star motif (motif 1) that follows from the conservation of the degree sequence, $d_i$:
\begin{equation}
\sum_{i=1}^{n}\left(
\begin{array}{c}
d_i\\
3\end{array}
\right)=N_1=n_1+n_4+2n_5+4n_6
\end{equation}
\begin{figure}[h]
\begin{center}
\includegraphics[scale=0.7]{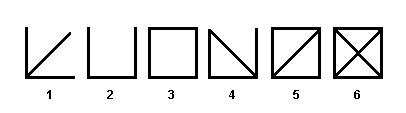}
\label{Fig.3}
\caption {The six 4 node motifs}
\end{center} 
\end{figure}

\section{Discussion}
The motif conservation laws show that in the case of directed 3 node motifs the induced subgraph statistics of the V-shaped motifs are completely determined by the statistics of the triangle shaped motifs. Consequently, the normalized triad significance of the 13 directed 3 node motifs has only 6 degrees of freedom due to the six conservation laws and the normalization which further reduces the degrees of freedom by one. Similarly, the subgraph ratio profile used in \cite{super} has only four degrees of freedom for motifs of size 4. The conservation laws further explain why the z-scores of triangle shaped motifs and V-shaped motifs are negatively correlated. 

The conservation laws could potentially be used to reduce the computational complexity of algorithms previously used to evaluate motifs since they show that counting of star shaped motifs is essentially redundant. 

The generalization of the conservation laws to higher order star shaped motifs and different edge and node types is straightforward. Moreover, when the counts of lower order motifs are conserved during the evaluation of higher order motifs \cite{motif} similar (approximate) conservation laws may arise. Recently, some generalizations of the configuration model that are based on specifying higher order subgraph degrees (such as triangle degree) for each node in addition to the edge degrees have been proposed \cite{newmanclus,karrer2010random}. In these generalized configuration models analogous conservation laws for higher order subgraphs that are not star shaped do also hold (again approximately) as a consequence of the conservation of the subgraph degrees. For instance, in the model proposed in \cite{newmanclus} the subgraph count of motif 4 in Fig.3 would be conserved approximately since such models contain only O(1) triangles in addition to those specified by the triangle degree.

\bibliography{citetations}
\bibliographystyle{unsrt}
\end{document}